# Strong Coupling Expansions for Antiferromagnetic Heisenberg $S = \frac{1}{2}$ Ladders

M. Reigrotzki*, H. Tsunetsugu*,†, and T.M. Rice*

*Theoretische Physik, ETH-Hönggerberg, 8093 Zürich, Switzerland

†Interdisziplinäres Projektzentrum für Supercomputing, ETH-Zentrum, 8092 Zürich, Switzerland

(July 1994)

## Abstract

The properties of antiferromagnetic Heisenberg $S = \frac{1}{2}$ ladders with 2, 3, and 4 chains are expanded in the ratio of the intra- and interchain coupling constants. A simple mapping procedure is introduced to relate the 4 and 2-chain ladders which holds down to moderate values of the expansion parameters. A second order calculation of the spin gap to the lowest triplet excitation in the 2- and 4-chain ladders is found to be quite accurate even at the isotropic point where the couplings are equal. Similar expansions and mapping procedures are presented for the 3-chain ladders which are in the same universality class as single chains.

PACS numbers:

Typeset using REVTEX

## I. INTRODUCTION

The study of antiferromagnetic Heisenberg spin-$\frac{1}{2}$ ladders has recently become of great interest because they offer the possibility of realizing spin liquid states with short range RVB (resonance valence bond) character [1–5]. Experimental systems currently being investigated are $(VO)_2P_2O_7$ [6,7] and the homologous series of compounds $Sr_{n-1}Cu_{n+1}O_{2n}$ [8,9]. The main evidence to date comes from numerical investigations using both Lanczos diagonalization of finite ladders [1,2,10], and density-matrix renormalization group [5], and quantum transfer-matrix methods [11]. These studies have shown that the low energy properties of ladders are governed by fixed point behavior determined by the limit of strong interchain and weak intrachain coupling. In this paper we present analytic expansions about this strong coupling limit for ladders with up to four legs (or chains). We compare these expansions with the numerical results and find that they hold qualitatively but not always quantitatively down to the isotropic limit (equal inter- and intrachain couplings). Further we introduce a mapping scheme to relate the 4-chain ladder to the simple (i.e. 2-chain) ladder and the 3-chain ladder to the single chain.

The Hamiltonian of such a ladder takes the form

$$H = \sum_{\leftrightarrow} J\, \mathbf{S}_l \cdot \mathbf{S}_m + \sum_{\updownarrow} J_\perp\, \mathbf{S}_l \cdot \mathbf{S}_m. \tag{1}$$

The intrachain coupling is $J$ and the interchain coupling across the rungs is $J_\perp$. The summation over the length of the chains — which will eventually tend to infinity — is denoted by $\leftrightarrow$ and the summation over the number of chains running parallel is denoted by $\updownarrow$.

In the strong coupling limit $J_\perp \gg J$, the part of the Hamiltonian with the sum *along* the chains is treated as a perturbation of the system of uncoupled rungs. For the simplest spin ladder (2-chains) the strong coupling perturbation expansion up to third order in $J/J_\perp$ is performed in Section 2. Ladders with three chains coupled together will be discussed in Section 3 and a mapping to the single Heisenberg spin-$\frac{1}{2}$ chain with an effective coupling determined by the splitting between the ground state and the first excited state of a $3 \times 2$ spin cluster. Lastly the case of 4-chain ladders is treated in Section 4 using two different approaches. First the four chains are considered as a system of two simple ladders running parallel, with a small coupling $J'_\perp$ between the two adjacent chains. Second, the four chains are treated in the same way as the simple ladder by starting in the limit of non-interacting rungs, each with four spins. In the second case — in analogy to the 3-chain system — a mapping to the simple ladder is performed.

## II. TWO COUPLED SPIN-$\frac{1}{2}$ CHAINS: THE SIMPLE SPIN LADDER

The best studied system is the simple spin-$\frac{1}{2}$ ladder, i.e. two parallel spin-$\frac{1}{2}$ chains strongly coupled together. The Hamiltonian has the form

$$H = H^0 + H^I, \qquad (2)$$

where $H^0$ and $H^I$ are given by

$$H^0 = \sum_{R=1}^{L} H^0_R = \sum_{R=1}^{L} J_\perp \left( \mathbf{S}^1_R \cdot \mathbf{S}^2_R \right) \qquad (3)$$

$$H^I = \sum_{R=1}^{L} H^I_R = \sum_{R=1}^{L} J \left( \mathbf{S}^1_R \cdot \mathbf{S}^1_{R+1} + \mathbf{S}^2_R \cdot \mathbf{S}^2_{R+1} \right). \qquad (4)$$

The intrachain (interchain) coupling is given by $J$ (respectively $J_\perp$), while $\mathbf{S}^i_R$ denotes the spin operator on the $R$-th rung on chain $i$. For chains of length $L$ periodic boundary conditions are introduced by defining $\mathbf{S}^i_{L+1} \equiv \mathbf{S}^i_1$.

The eigenstates of the rung Hamiltonian $H^0_R$ are given by a singlet state $|s\rangle$ with energy $E_s = -\frac{3}{4} J_\perp$ and three triplet states $|t^\sigma\rangle$ with spin z-component $\sigma = -1, 0, 1$ and energy $E_t = \frac{1}{4} J_\perp$. The eigenstates of $H^0$ are direct products of rung states.

### A. Spin Gap and Dispersion relation of magnon excitations

When applying perturbation theory in $J/J_\perp$ for the strong-coupling limit, the first order correction to the ground state energy vanishes, since $H^I$ pairwise excites two adjacent singlets of the unperturbed ground state $|0\rangle = |s \ldots s\rangle$ to a linear combination of triplets with total spin $S = 0$:

$$H^I |0\rangle = \frac{1}{2} J \sum_R \left| s \ldots (t^0 t^0 - t^+ t^- - t^- t^+) \ldots s \right\rangle. \qquad (5)$$

This excitation leads to a correction of the ground state energy. Up to third order perturbation in the strong coupling limit, the ground-state energy per rung is

$$\frac{E_g}{L} = -J_\perp \left[ \frac{3}{4} + \frac{3}{8} \left( \frac{J}{J_\perp} \right)^2 + \frac{3}{16} \left( \frac{J}{J_\perp} \right)^3 \right]. \qquad (6)$$

The first excited state of the unperturbed ladder is obtained by promoting one rung to a triplet state. The $L$-fold degeneracy of this state is lifted in first order perturbation in the strong-coupling limit. To first order in $J/J_\perp$ the eigenstates are given by Bloch states

$$|1^\sigma, k\rangle = \frac{1}{\sqrt{L}} \sum_{R=1}^{L} e^{ikx_l} |s \ldots t^\sigma_l \ldots s\rangle, \qquad (7)$$

where the $l$-th rung is excited to a triplet with $S^z = \sigma$. The excitation energy $\omega(k)$ for magnon excitations up to third order $J/J_\perp$ in the strong-coupling limit is

$$\frac{\omega(k)}{J_\perp} = 1 + \left( \frac{J}{J_\perp} \right) \cos k + \frac{1}{4} \left( \frac{J}{J_\perp} \right)^2 (3 - \cos 2k) \qquad (8)$$
$$- \frac{1}{8} \left( \frac{J}{J_\perp} \right)^3 (2 \cos k + 2 \cos 2k - \cos 3k - 3).$$

To second order this result is equal to that obtained by Barnes et. al. in [2], except for the additional second order term $\propto \cos 2k$. The energy has a minimum at $k = \pi$. Therefore the spin gap $\Delta = \omega(\pi)$ is

$$\frac{\Delta}{J_\perp} = 1 - \left( \frac{J}{J_\perp} \right) + \frac{1}{2} \left( \frac{J}{J_\perp} \right)^2 + \frac{1}{4} \left( \frac{J}{J_\perp} \right)^3. \qquad (9)$$

In Fig. 1 the dispersion relation in second and third order is plotted for a ratio of $J/J_\perp = 0.5$. The third order correction improves the agreement to numerical data from [2]. The deviation near $k = 0$ is due to 2 magnon processes. Fig. 2 shows the spin gap in second and third order compared to numerical data for a $2 \times 8$ Heisenberg ladder [1]. At the isotropic point $J/J_\perp = 1$ results from Barnes et. al. [2] for $2 \times 8$ and $2 \times \infty$ ladders are included. In the isotropic region, the third order correction unfortunately leads to worse agreement than the second order correction, which is surprisingly good in the infinite-length limit for the simple Heisenberg spin ladder.

## III. LADDERS WITH THREE SPIN-$\frac{1}{2}$ CHAINS

A system of three coupled spin-$\frac{1}{2}$ chains has a degenerate ground state in the strong-coupling limit which leads to an additional complication compared to the study of the simple ladder (or any other system of an even number of chains coupled together). White et. al. [5] give an explanation of the fundamental difference of even and odd number of spin chains coupled together as due to the behavior of topological spin defects. Since the rung states are already degenerate, perturbation theory for degenerate systems must be used from the beginning. This can be realized by mapping the system to the single Heisenberg spin-$\frac{1}{2}$ chain — which has been studied extensively — using an effective coupling in first order perturbation in $J/J_\perp$. Later we use the exact diagonalization of a $3 \times 2$ spin cluster to determine the effective coupling.

The Hamiltonian of the system is given in analogy to the one of the ladder by

$$H = H^0 + H^I, \tag{10}$$

with

$$H^0 = \sum_{R=1}^{L} H_R^0 = \sum_{R=1}^{L} J_\perp \left( \mathbf{S}_R^1 \cdot \mathbf{S}_R^2 + \mathbf{S}_R^2 \cdot \mathbf{S}_R^3 \right) \tag{11}$$

$$H^I = \sum_{R=1}^{L} H_R^I = \sum_{R=1}^{L} \sum_{j=1}^{3} J\, \mathbf{S}_R^j \cdot \mathbf{S}_{R+1}^j. \tag{12}$$

The energy levels of a rung system together with their degeneracy are depicted in Fig. 3a. The two spin-$\frac{1}{2}$ states are denoted by $|d^\sigma\rangle$ (doublet), $\sigma$ corresponding to $S^z = \pm\frac{1}{2}$, the spin-$\frac{3}{2}$ state by $|q^{\tilde\sigma}\rangle$ (quartet) with $\tilde\sigma$ corresponding to $S^z = -\frac{3}{2}, -\frac{1}{2}, \frac{1}{2}, \frac{3}{2}$.

The energy levels of an uncoupled pair of rungs (with Hamiltonian $H^0 = H_{R=1}^0 + H_{R=2}^0$) and their splitting into spin-subspaces upon switching on the perturbation $H^I = H_{R=1}^I$ are shown in Fig. 3b.

In first order perturbation theory the ground state of a pair of rungs is

$$E_g = -2J_\perp - \tfrac{3}{4}J \tag{13}$$

and the splitting to the lowest triplet is equal to $J$. The effective coupling in first order perturbation theory in the strong coupling limit then is

$$J_{\text{eff}} \equiv J. \tag{14}$$

The results of the exact diagonalization of a $3 \times 2$ cluster are shown in Fig. 4. The separation of the second from the first excited state is greater than the splitting $J_{\text{eff}}$ of the singlet and the triplet, reaching its minimum of $\cong J_{\text{eff}}$ at the isotropic point $J/J_\perp = 1$. Thus the mapping should give reasonable results for temperatures $k_B T \leq J_{\text{eff}}$. At isotropy, from Fig. 5 we obtain a value $J_{\text{eff}} = 0.82 J$. A more complete density matrix renormalization group (DMRG) study of isotropic Heisenberg coupled chains by White et. al. [5] leads to qualitatively comparable results. They consider finite ladders with open boundary conditions so that there is a finite excitation energy determined by the velocity of the des Cloizeaux-Pearson mode. In a single chain this is proportional to the coupling constant. The ratio of these velocities in the 3-chain and 1-chain systems gives a direct measure of renormalized coupling $J_{\text{eff}} \approx 0.68$.

## IV. LADDERS WITH FOUR COUPLED SPIN-$\frac{1}{2}$ CHAINS

To study the ladders with 4 parallel coupled Heisenberg spin-$\frac{1}{2}$ chains in the strong coupling limit, two different approaches are taken. First the 4 chains will be treated as two simple ladders coupled together and expanded around the strong rung coupling limit. Secondly, we map the 4-chain system to a simple double-chain ladder with renormalized coupling constants. The Hamiltonian of this system is the sum of three terms $H = H^0 + H^1 + H^2$ with

$$H^0 = \sum_{R=1}^{L} J_\perp \left( \mathbf{S}_R^1 \cdot \mathbf{S}_R^2 + \mathbf{S}_R^3 \cdot \mathbf{S}_R^4 \right) \tag{15}$$

$$H^1 = \sum_{R=1}^{L} \sum_{j=1}^{4} J\, \mathbf{S}_R^j \cdot \mathbf{S}_{R+1}^j \tag{16}$$

$$H^2 = \sum_{R=1}^{L} J_\perp'\, \mathbf{S}_R^2 \cdot \mathbf{S}_R^3. \tag{17}$$

As in the double chain system, the interchain coupling (coupling between chains one and two and between three and four) is $J_\perp$ and the intrachain coupling is $J$. Periodic boundary conditions again are introduced by $\mathbf{S}_{L+1}^i \equiv \mathbf{S}_1^i$. The coupling *between the two ladders is $J'_\perp$.*

### A. Two coupled spin-$\frac{1}{2}$ ladders

The part $H^0 + H^1$ of $H$ is the sum of two simple ladder Hamiltonians:

$$H^0 + H^1 \equiv H_u + H_l. \tag{18}$$

The indices $u$ and $l$ denote simple ladder Hamiltonians (2) on the *upper* resp. *lower* simple ladder.

To determine the ground state energy and the energy of the low-lying excitations, perturbation in $J$ and $J'_\perp$ in the strong coupling limit $J \approx J'_\perp \ll J_\perp$ up to second order is applied. The first excited state is $2 \cdot 3L$-fold degenerate, the factor 2 arising since promoting a singlet to a triplet on one rung can be done on either of the two coupled ladders. The $L$-fold degeneracy is lifted as in the simple ladder system by the perturbation $H^1$ (transforming to Bloch states); the remaining 3-fold degeneracy is a spin degeneracy. The second perturbing Hamiltonian $H^2$ will lift the 2-fold degeneracy into even and odd parity states, the parity-transformation $P_{\text{chain}}$ being defined by reversing the chain order ($P_{\text{chain}} : i \to 5 - i$).

For the ground state the energy per rung to second order in $J$ and $J'_\perp$ is

$$\frac{E_g}{L} = -J_\perp \left[\frac{3}{2} + \frac{3}{4}\left(\frac{J}{J_\perp}\right)^2 + \frac{3}{32}\left(\frac{J'_\perp}{J_\perp}\right)^2\right]. \tag{19}$$

With the corrections obtained for the first excited state with odd ($-$) and even parity ($+$), the dispersion relation for the 4-chain system as two coupled ladders up to second order perturbation in the strong coupling limit is

$$\frac{\omega^\mp(k)}{J_\perp} = 1 + \left(\frac{J}{J_\perp}\right)\cos k + \frac{1}{4}\left(\frac{J}{J_\perp}\right)^2(3 - \cos 2k) - \frac{1}{32}\left(\frac{J'_\perp}{J_\perp}\right)^2 \tag{20}$$

$$\mp \left[\frac{1}{4}\left(\frac{J'_\perp}{J_\perp}\right) + \frac{1}{8}\left(\frac{J'_\perp}{J_\perp}\right)^2 - \frac{1}{4}\frac{J'_\perp J}{J_\perp^2}\cos k\right]$$

The dispersion relation for the two branches is shown in Fig. 6 for $J = J'_\perp = \frac{1}{2}J_\perp$. Fig. 7 shows the spin gap for the two branches.

### B. Four coupled spin-$\frac{1}{2}$ chains

The Hamiltonian is again of the form (15)–(17). The interaction Hamiltonian $H^2$ between ladders will not be treated as perturbation of the one-rung eigenstates of the double ladder but instead exact eigenstates of the sum $H^0 + H^2$ are the basis-states for a perturbative treatment of the intrachain coupling $J$. Additionally the problem is mapped to the simple ladder by exact diagonalization of the $4 \times 2$ cluster.

#### 1. Exact one-rung eigenstates

The 16 one-rung eigenstates of $H^0 + H^2$ are denoted by

$$\begin{array}{ll}
|s_\mp\rangle & E_{s_\mp} = -\frac{1}{4}(2J_\perp + J'_\perp) \mp \frac{1}{2}\sqrt{4J_\perp^2 - 2J_\perp J'_\perp + J'^2_\perp} \\
|t_\alpha^\sigma\rangle & E_{t_\alpha} = -\frac{1}{4}J'_\perp - \frac{1}{2}\sqrt{J_\perp^2 + J'^2_\perp} \\
|t_\beta^\sigma\rangle & E_{t_\beta} = -\frac{1}{4}(2J_\perp - J'_\perp) \\
|t_\gamma^\sigma\rangle & E_{t_\gamma} = -\frac{1}{4}J'_\perp + \frac{1}{2}\sqrt{J_\perp^2 + J'^2_\perp} \\
|q^{\tilde\sigma}\rangle & E_q = \frac{1}{4}(2J_\perp + J'_\perp)
\end{array} \tag{21}$$

There are two singlets $|s_\mp\rangle$, three triplets $|t_\alpha^\sigma\rangle$, $|t_\beta^\sigma\rangle$ and $|t_\gamma^\sigma\rangle$ with $S^z$ components $\sigma = -1, 0, 1$ and a quintet $|q^{\tilde\sigma}\rangle$ with $S^z$ components $\tilde\sigma = -2, -1, 0, 1, 2$.

For $J'_\perp = J_\perp$, the ground state energy per spin of the 4-chain system is

$$\frac{E_g}{4L} = -J_\perp \left[\frac{3 + 2\sqrt{3}}{4} + \frac{16 + 3\sqrt{3}}{24}\left(\frac{J}{J_\perp}\right)^2\right], \tag{22}$$

up to the second order in $J/J_\perp$. By the same reasoning as for the double ladder, the coupling $J'_\perp$ leads to a splitting of the first excitation from the ground state into an even and odd parity state. For the odd parity state the dispersion relation for the low-lying excitations of four coupled Heisenberg spin-$\frac{1}{2}$ chains up to second order perturbation theory in the strong coupling limit is

$$\frac{\omega^-(k)}{J_\perp} = 0.659 + 1.075 \left(\frac{J}{J_\perp}\right) \cos k$$
$$+ \left(\frac{J}{J_\perp}\right)^2 (1.086 - 0.035 \cos k - 0.469 \cos 2k), \qquad (23)$$

and

$$\frac{\omega^+(k)}{J_\perp} = 1.366 + 0.667 \left(\frac{J}{J_\perp}\right) \cos k$$
$$+ \left(\frac{J}{J_\perp}\right)^2 (1.826 - 1.155 \cos k - 0.081 \cos 2k) \qquad (24)$$

for the even parity state. The dispersion relation for the two branches is shown in Fig. 8. For the odd parity branch, the spin gap minimum is always at $k = \pi$. On the other hand, the minimum of the even parity branch jumps from $k = \pi$ to $k = 0$ at $J/J_\perp = 0.6$. This explains the kink in the curve of the even parity spin gap in Fig. 9.

### C. Mapping to an effective simple ladder

As in Section 3, let us map the 4-chain ladder into an effective simple ladder to determine the spin gap. Throughout this section, we assume $J'_\perp = J_\perp$ again. To this end, we calculate the eigenenergies of a $4 \times 2$ cluster, and determine the effective coupling constants of the effective simple ladder, $J^* = J^*(J, J_\perp)$, $J^*_\perp = J^*_\perp(J, J_\perp)$, so as to reproduce the same low-energy spectrum. In this approximation, the spin gap of the 4-chain ladder is then given by that of the simple ladder as

$$\Delta_{\text{4-chain}}(J, J_\perp) = \Delta_{\text{2-chain}}(J^*(J, J_\perp), J^*_\perp(J, J_\perp)). \qquad (25)$$

The energy spectrum of a $2 \times 2$ cluster is easily calculated. The ground state is a spin singlet, and there are two magnons, bonding and antibonding combinations of the two states in which one rung is a spin singlet and the other is a triplet. The energies of these three lowest states are

$$E_0 = -\tfrac{1}{2}(J^* + J^*_\perp) - \sqrt{J^{*2} - J^* J^*_\perp + J^{*2}_\perp}, \qquad (26)$$
$$E_{1,\pm} = -\tfrac{1}{2} J^*_\perp \pm \tfrac{1}{2} J^*. \qquad (27)$$

These three states are sufficient to determine the coupling constants:

$$J^* = E_{1,+} - E_{1,-} , \qquad (28)$$
$$J^*_\perp = \sqrt{D^2 - DJ^* - \tfrac{1}{2}J^{*2}} + \tfrac{1}{2}J^*, \qquad (29)$$

where $D \equiv \tfrac{1}{2}(E_{1,+} + E_{1,-}) - E_0$ is the average energy separation between the ground state singlet and the two triplets. There is one other spin triplet consisting of two magnons, with the energy, $E_{1,tt} = \tfrac{1}{2}(J^*_\perp - J^*)$.

Fig. 10 shows the energy levels of the $4 \times 2$ cluster as a function of the intrachain coupling, $J$. The ground state is a spin singlet and there are two triplets above it. The two triplets are split with a separation increasing with $J$. This splitting corresponds to the band width of propagating magnons along the chain direction. At around $J/J_\perp \sim 0.64$, another triplet goes down and becomes lower than $E_{1,+}$. This triplet is the two magnon state. Therefore, this crossing corresponds to the point where the bottom of the two-magnon continuum becomes lower than the top of the one-magnon mode. However, these two triplet eigenstates, having different parities with respect to the mirror symmetry, $P_{\text{chain}}$, do not mix to each other, and this level crossing does not have significant consequences.

The energy spectrum Fig. 10 has the same structure as the $2 \times 2$ system, at least for small $J$'s. The corresponding effective couplings are determined by (28) and (29). The result is shown in Fig. 11. The mapping breaks down for large couplings, $J/J_\perp > 0.70$, where the condition $D \geq \frac{\sqrt{3}+1}{2}(E_{1,+} - E_{1,-})$, which is necessary for a real $J^*_\perp$, is no longer satisfied.

We can now estimate the spin gap of the 4-chain ladder by using (25). Here we use a Padé approximation for the $\Delta_{\text{2-chain}}$ determined by the strong coupling limit (9) and the weak-coupling asymptotic form, $\Delta \propto J_\perp$ [12]:

$$\Delta_{\text{2-chain}}(J, J_\perp) = J_\perp G\left(\frac{J_\perp}{J}\right), \qquad (30)$$
$$G(x) = \frac{(2+2a)x^2 + (1-3a)x + 2a}{(2+2a)x^2 + (3-a)x + 2}, \qquad (31)$$

where $a \equiv \lim_{J_\perp \to 0} \Delta_{\text{2-chain}}/J_\perp$. This Padé approximant has a correct asymptotic form in both limits, $J_\perp/J \to \infty$ and $J_\perp/J \to 0$. We use here $a = 1$ determined by $\Delta_{\text{2-chain}} = 0.5$ at $J_\perp = J$, but this form does not agree so well with the numerical results [2] at $0 < J_\perp/J < 1$.

Figure 12 shows the result of the spin gap for the 4-chain ladder. The gap is obtained by using (31) for (25) with the values of $J^*$'s and $J_\perp^*$'s shown in Fig. 11. The values determined by numerical diagonalization [5,10] are included at $J_\perp = J$. The result of the second-order perturbation (23) is also included. This mapping shows a correct tendency.

## V. CONCLUSIONS

In this paper we presented the results of analytic expansions about the limit of strong interchain coupling. It is convenient to expand around this strong coupling limit since the qualitative behavior and therefore the universality class of the system remains unchanged. An analytic expansion allows one to examine the evolution of the system to the case of isotropic coupling. A magnon is an elementary $S = 1$ excitation, but the spin density distribution which is localized on a single rung in the strong coupling limit evolves continuously into a distribution spread over several rungs at isotropy. With decreasing interchain coupling the magnon spectrum changes its dispersion relation from a simple cosine band as the size of the magnons increases and longer range hopping matrix elements enter. An $S = 1$ magnon can be regarded as a bound state of two $S = \frac{1}{2}$ spinons [13] on individual chains, but since the interchain coupling changes the excitation spectrum completely this analogy is only qualitative at best. The 4-chain ladder also has a spin gap to $S = 1$ magnon excitations and so belongs to the same universality class as the 2-chain ladder, and a mapping procedure between 4-chain and 2-chain ladders is possible. This works well down to moderate values of the intra- to interchain coupling but not down to the isotropic limit where they are equal. On the other hand a second order perturbation for the spin gap works surprisingly well down to the isotropic limit. The 3-chain ladder can be mapped onto the single chain and a simple mapping procedure is found to work quite well down to the isotropic limit when compared to numerical results. The low energy behavior of the two systems will be similar and the spinons will now extend over the three chains. Lastly the success of the expansion around the strong coupling limit of the Heisenberg ladders encourages one to consider a corresponding expansion for the doped ladders described by a $t$-$J$ model which have shown interesting results in mean field and numerical investigations [14,15].

FIGURES

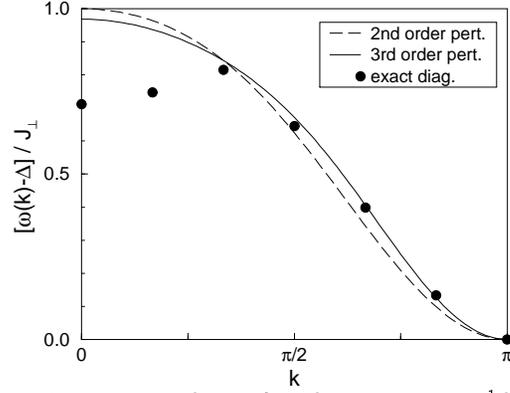

FIG. 1. Dispersion relation of simple Heisenberg spin-$\frac{1}{2}$ ladder for $J/J_\perp = 0.5$. The data of exact diagonalization are taken from Ref. [2].

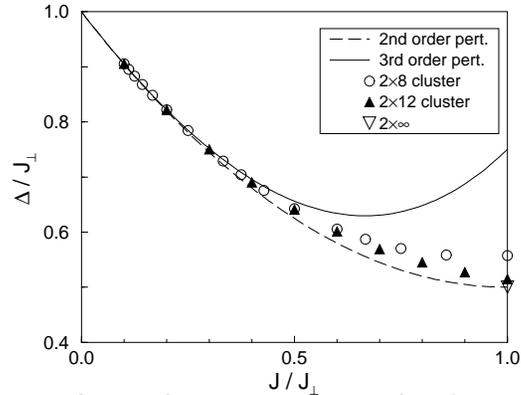

FIG. 2. Spin gap for a magnon excitation of simple Heisenberg spin-$\frac{1}{2}$ ladder. The extrapolated value for $2 \times \infty$ is taken from Ref. [2]

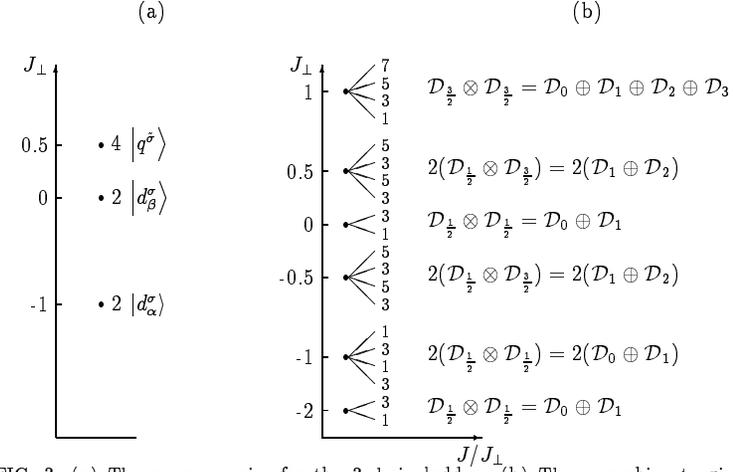

FIG. 3. (a) The rung energies for the 3-chain ladder. (b) These combine to give the eigenenergies of a pair of rungs, and under the action of $H^I$ they split yielding the spectrum for a $3 \times 2$ cluster.

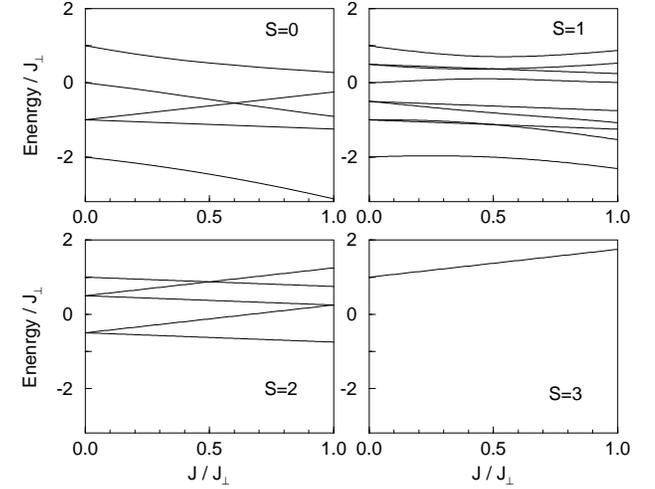

FIG. 4. Diagonalization of $3 \times 2$ cluster classified by total spin $S$.

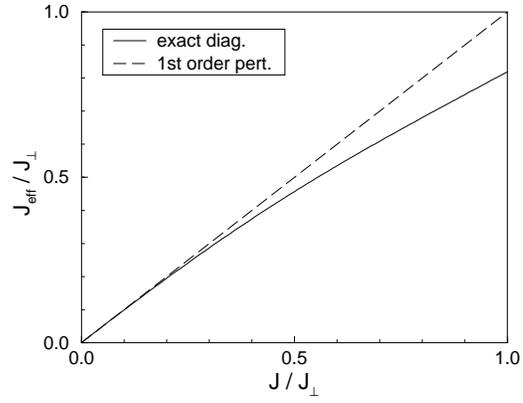

FIG. 5. Effective coupling obtained by mapping the 3-chain to the simple chain.

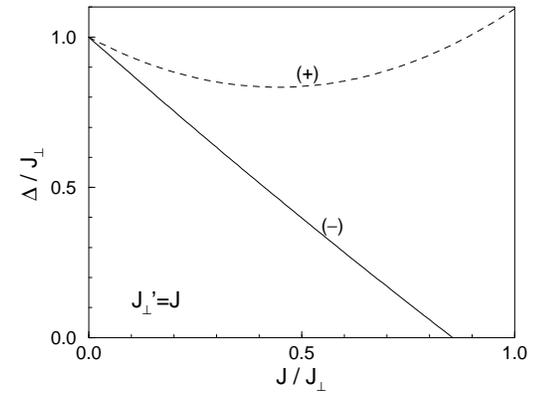

FIG. 7. Spin gap for a magnon of double Heisenberg spin-$\frac{1}{2}$ ladder.

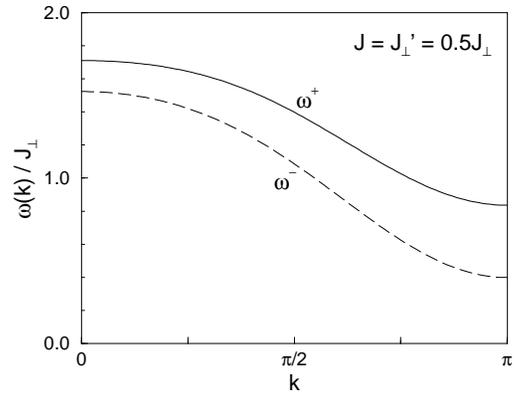

FIG. 6. Dispersion relation for double Heisenberg spin-$\frac{1}{2}$ ladder.

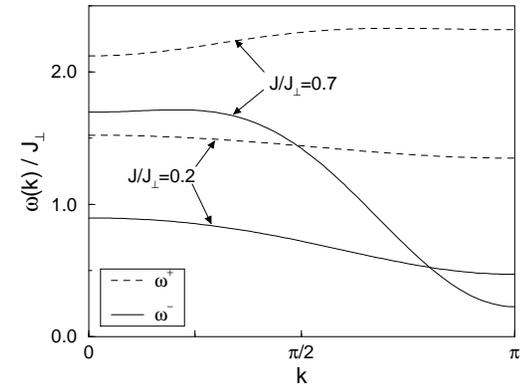

FIG. 8. Dispersion relation for four coupled Heisenberg spin-$\frac{1}{2}$ chains.

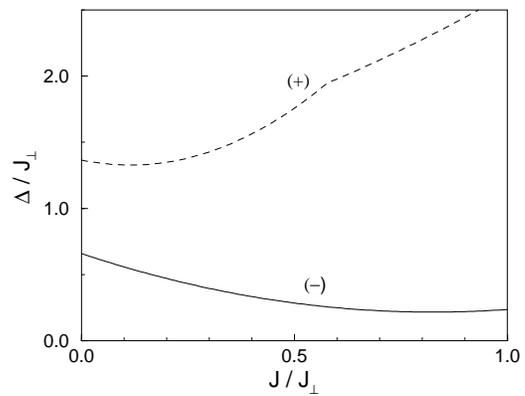

FIG. 9. Spin gap for a magnon of four Heisenberg spin-$\frac{1}{2}$ chains.

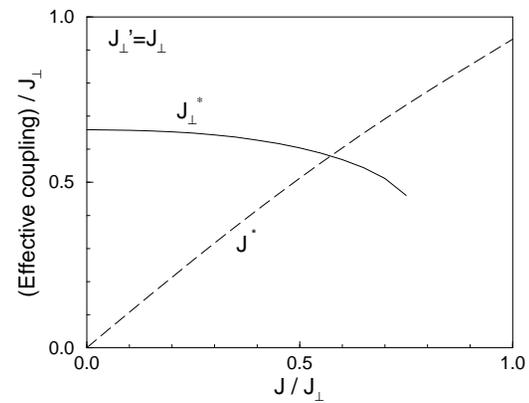

FIG. 11. Effective couplings when the 4-chain ladder is mapped to the simple ladder.

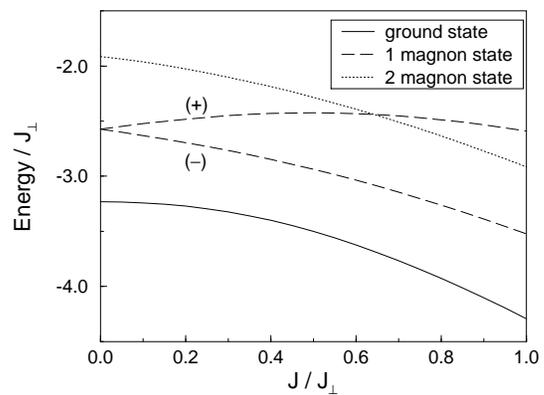

FIG. 10. Energies of $4 \times 2$ cluster.

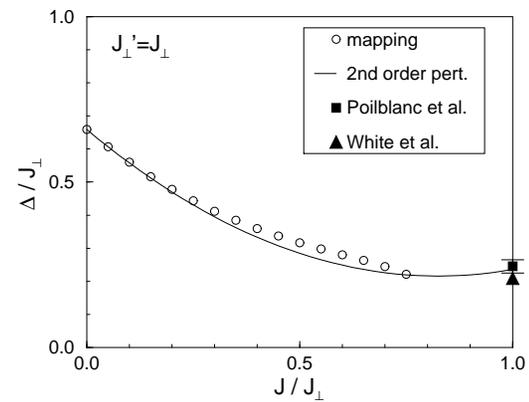

FIG. 12. Spin gap of the 4-chain ladder calculated by mapping to a simple ladder.